\pdfoutput=1

\documentclass[11pt]{article}
\usepackage{hyperref} 

\usepackage{algorithm}
\usepackage{algpseudocode}

\usepackage[final]{acl}

\usepackage{times}
\usepackage{latexsym}

\usepackage{amsmath}
\usepackage{amssymb}
\usepackage{booktabs} 
\usepackage{multirow}
\usepackage{graphicx}
\usepackage{xcolor}
\usepackage{colortbl} 
\usepackage{makecell} 
\definecolor{defgreen}{RGB}{235, 250, 235}

\usepackage[T1]{fontenc}

\usepackage[utf8]{inputenc}

\usepackage{microtype}

\usepackage{inconsolata}

\usepackage{graphicx}

\definecolor{warningcolor}{RGB}{255,97,0}
\title{JPU: Bridging Jailbreak Defense and Unlearning via On-Policy Path Rectification
\\ {\color{warningcolor} \normalsize Warning: This paper contains potentially harmful LLMs-generated content.}}

\author{
\textbf{Xi Wang}\textsuperscript{$\clubsuit$} \;\;\; 
\textbf{Songlei Jian}\textsuperscript{$\clubsuit$ $^{\dagger}$} \;\;\;
\textbf{Shasha Li}\textsuperscript{$\clubsuit$ $^{\dagger}$} \;\;\; 
\textbf{Xiaopeng Li}\textsuperscript{$\clubsuit$} \;\;\; \\
\textbf{Zhaoye Li}\textsuperscript{$\clubsuit$} \;\;\; 
\textbf{Bin Ji}\textsuperscript{$\clubsuit$} \;\;\; 
\textbf{Baosheng Wang}\textsuperscript{$\clubsuit$} \;\;\; 
\textbf{Jie Yu}\textsuperscript{$\clubsuit$ $^{\dagger}$} \\
  \textsuperscript{$\clubsuit$}National University of Defense Technology\; \\
  \texttt{\{wx\_23ndt,jiansonglei,shashali,xiaopengli\}@nudt.edu.cn}\\ 
  \texttt{\{lizhaoye23, jibin, bswang,yj\}@nudt.edu.cn}
}

\begin{document}
\maketitle
\begingroup
\endgroup
\begin{abstract}
Despite extensive safety alignment, Large Language Models (LLMs) often fail against jailbreak attacks. While machine unlearning has emerged as a promising defense by erasing specific harmful parameters, current methods remain vulnerable to diverse jailbreaks. We first conduct an empirical study and discover that this failure mechanism is caused by jailbreaks primarily activating non-erased parameters in the intermediate layers. Further, by probing the underlying mechanism through which these circumvented parameters reassemble into the prohibited output, we verify the persistent existence of dynamic \textbf{jailbreak paths} and show that the inability to rectify them constitutes the fundamental gap in existing unlearning defenses. To bridge this gap, we propose \textbf{J}ailbreak \textbf{P}ath \textbf{U}nlearning (JPU), which is the first to rectify dynamic jailbreak paths towards safety anchors by dynamically mining on-policy adversarial samples to expose vulnerabilities and identify jailbreak paths. Extensive experiments demonstrate that JPU significantly enhances jailbreak resistance against dynamic attacks while preserving the model's utility.
\end{abstract}

\section{Introduction}

\begin{figure}[!ht]
\centering
\includegraphics[width=\linewidth]{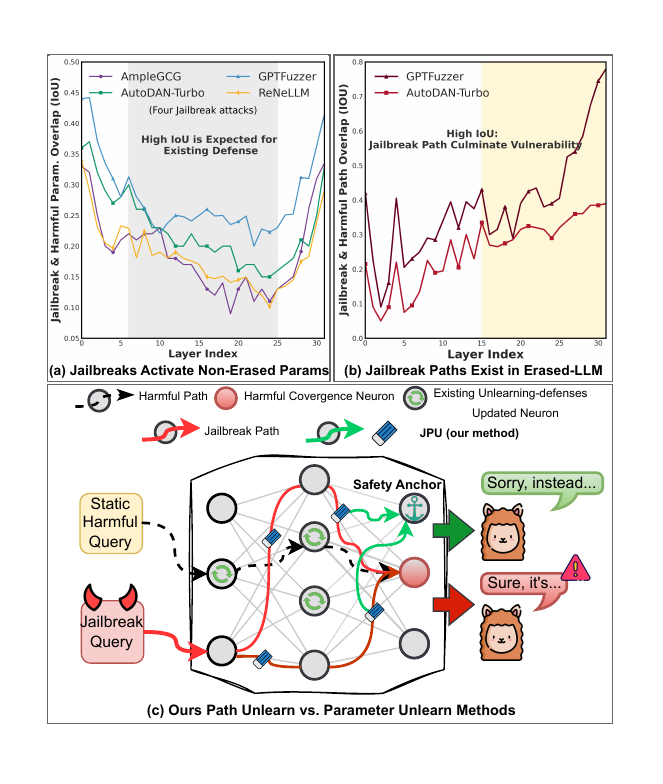} 
\caption{An illustration of how jailbreak path unlearning enhances jailbreak defense performance. Compared to existing unlearning defenses that exhibits the failure mechanism (a), our proposed method JPU (c) explicitly targets and rectifies the underlying issue, the existing jailbreak paths within erased-LLMs (b), achieving superior defense performance while preserve utility.}
\label{ana}
\end{figure}

The rapid evolution of Large Language Models (LLMs), exemplified by ChatGPT~\cite{OpenAI}, Gemini~\cite{team2023gemini}, and DeepSeek~\cite{guo2025deepseek}, has contributed significantly to the rise of Artificial Intelligence. These models demonstrate exceptional performance across complex tasks, including content generation, code synthesis, and mathematical reasoning~\cite{ahn2024large, hu2024dynamic}. However, the deployment of LLMs exposes significant safety vulnerabilities that pose substantial risks to modern society~\cite{nadeem2021stereoset, phanireddy2025llm}. To alleviate these risks, significant efforts have focused on advanced alignment strategies to construct safety-aligned LLMs~\cite{kaneko2022debiasing, zhao2024improving}. 

Despite these efforts, recent studies reveal that aligned LLMs remain brittle against ``Jailbreak Attacks''~\cite{goldstein2023generative, chu2024comprehensive}, which manipulate models into generating prohibited content. To mitigate these risks, researchers have developed various defenses, including decoding enhancement~\cite{xu2024safedecoding}, model editing~\cite{zhao2024defending}, and machine unlearning~\cite{lu2024eraser, shi2025safety}. Among these, machine unlearning has emerged as a promising defense paradigm aiming to directly remove specific harmful knowledge from the model parameters and retraining LLMs to reject malicious requests.
Existing unlearning defenses, which primarily focus on erasing isolated parameters and static datasets, remain vulnerable to evolving jailbreak attacks. To investigate this failure, we analyze the overlap of activated parameters between successful jailbreak queries and standard harmful queries on an erased LLM~\cite{lu2024eraser} using SNIP importance analysis~\cite{wei2024assessing},  which identifies the most salient neurons driving specific behaviors. As illustrated in Figure~\ref{ana}(a), parameters activated by diverse attacks in intermediate layers exhibit minimal overlap with the ``erased'' regions associated with harmful queries, contradicting the high overlap expected for existing unlearning defenses. This suggests that jailbreak behaviors are not strictly bound to a fixed set of isolated neurons.

While analyzing isolated parameters is a common practice, it fails to reveal the underlying attack mechanism by overlooking the inter-layer connectivity essential for orchestrating distributed activations toward harmful outputs. To address this gap, we hypothesize that dynamic \textbf{jailbreak paths} serve as the determinant factor, routing information through non-erased parameters across multiple layers to successfully generate unsafe content. We verify this by tracing the alignment between jailbreak-induced paths and direct-harm trajectories using Inter-layer Gradient Integration (IGI)~\cite{li2025cross}.  As illustrated in Figure~\ref{ana}(b), diverse jailbreaks progressively converge with the harmful pathways from intermediate to final layers, culminating in the unsafe outputs. This confirms that dynamic jailbreak paths in supposedly ``unlearned'' models underlie persistent vulnerabilities, exposing a fundamental defense gap: current static parameter unlearning defenses are insufficient to eliminate dynamic path-based attack mechanisms, leaving models susceptible to adaptive jailbreaks.

To bridge this gap, we propose \textbf{J}ailbreak \textbf{P}ath \textbf{U}nlearning (JPU) Figure~\ref{ana}(c), a unified unlearning framework designed to rectify dynamic jailbreak paths towards safety anchors by dynamically mining on-policy adversarial samples to expose vulnerabilities and identify jailbreak paths. JPU comprises three key components: on-policy attack buffer mining, jailbreak path identification, and constrained path rectification. Specifically, JPU first mines vulnerable adversarial samples from an adaptive attack buffer by estimating the current model’s refusal tendency. It then retraces critical jailbreak paths flowing from intermediate layers to harmful sink nodes in deep layers using a variant of inter-layer gradient integration. Finally, JPU performs constrained jailbreak path rectification to align the retraced paths with safety anchors and desired behaviors. Meanwhile, a utility preservation constraint is incorporated to maintain the model’s general capabilities. Experimental results demonstrate that JPU achieves superior jailbreak defense performance while preserving utility compared with existing unlearning defense methods.


In summary, our contributions are as follows:
\begin{itemize}
    \item We conduct progressive empirical analysis to expose the failure mechanisms and the critical gap of existing unlearning defenses: jailbreak activation of non-erased parameters and dynamic jailbreak paths exist in erased LLMs.
    \item We introduce JPU, the first unlearning framework that rectifies dynamic jailbreak paths toward safety anchors by dynamically mining on-policy adversarial samples to expose vulnerabilities and identify jailbreak paths.
    \item Extensive experiments demonstrate that JPU achieves superior jailbreak defense performance compared to existing methods, while effectively preserving general utility.
\end{itemize}

\section{Related Work}
\subsection{Jailbreak Attack and Defense}
While Large Language Models are aligned with human values via safety training techniques~\cite{wang2024comprehensive, dong2024rlhf}, recent studies reveal that these safeguards remain brittle. Adversarial users can exploit vulnerabilities through jailbreak attacks to elicit prohibited content. 
Existing attacks can generally be categorized by access privileges. White-box attacks leverage accessible model gradients and internal states to optimize adversarial suffixes or manipulate decoding distributions~\cite{zhang2024jailbreak, zou2023universal}. Conversely, black-box attacks rely on designing complex prompt templates, genetic algorithms, or iterative refinement via attacker LLMs to probe safety boundaries~\cite{yu2023gptfuzzer, chao2025jailbreaking}.

Defense strategies have evolved in response to these threats, primarily falling into two categories: input/output filtering and internal alignment. Filtering methods focus on detecting harmful patterns but often incur significant inference latency and are prone to false positives~\cite{lirain, robeysmoothllm}. Internal methods attempt to adjust model parameters or decoding strategies to steer generation towards safety~\cite{li2025detam, xie2024gradsafe}. However, most existing internal defenses are static, as they lack the dynamic adaptability required to counter the evolving diversity of jailbreak patterns, leaving vulnerability to adaptive attacks.

\subsection{Unlearning against LLM Jailbreak}
\label{unlearn_base}
The harmful capabilities of LLMs source from inevitably contained toxic content within vast pre-training corpora. Machine Unlearning is the methodology which aims to surgically erase specific harmful knowledge or capabilities to offer a promising avenue for safety enforcement. Current unlearning defenses strive to balance safety erasure with utility preservation. These methods can be broadly classified into global unlearning and selective unlearning. Global methods update all model parameters to maximize the loss on harmful data~\cite{lu2024eraser,zhang2024safe}. Selective methods  attempt to first localize safety-critical parameters and then apply targeted updates to minimize side effects~\cite{shi2025safety,jia2024wagle}. 

Despite progress, our empirical analysis uncovers a critical oversight in this paradigm: erased-LLMs retain active jailbreak paths, leaving significant vulnerabilities to adversarial manipulation. In contrast, our approach fill this gap by explicitly identifying and unlearning the dynamic jailbreak paths exposed by diverse attacks.


\begin{figure*}[!ht]
\centering
\includegraphics[width=1\linewidth]{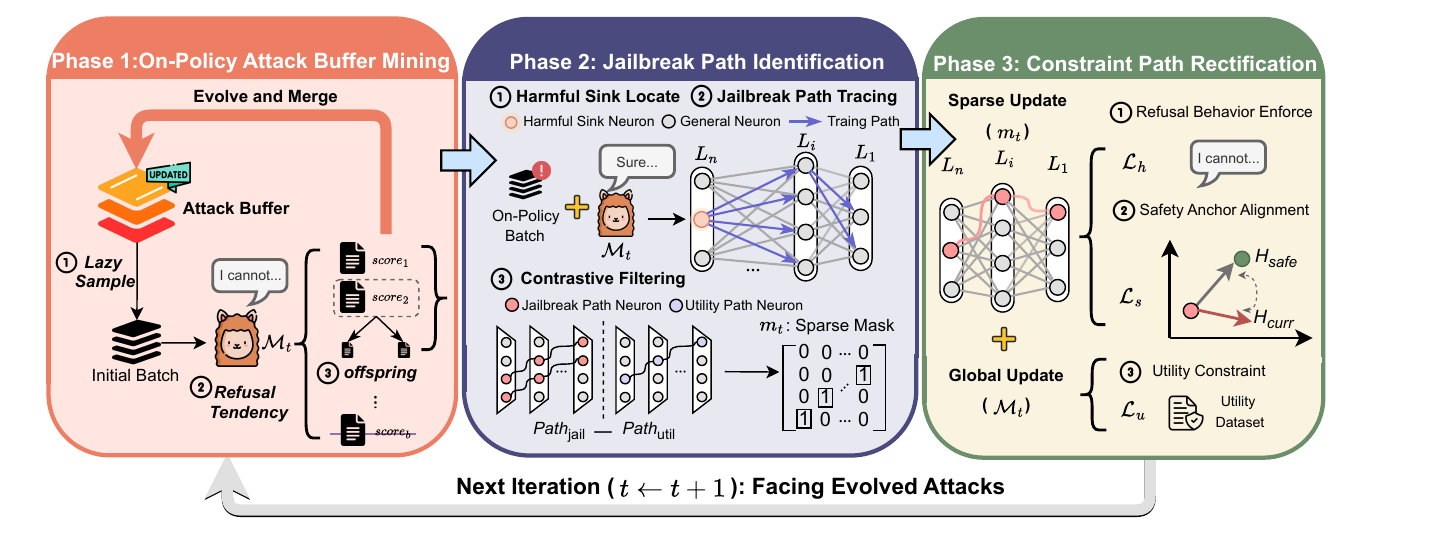} 
\caption{Overview of JPU. It consists of three steps. \textbf{On-Policy Attack Buffer Mining:} We mine adversarial samples from current model to form on-policy vulnerable attack batch. \textbf{Jailbreak Path Identification:} We trace critical jailbreak paths flowing from intermediate layers to harmful sink nodes in deep layers. \textbf{Constraint Path Rectification:} Under three constraints, we perform path rectification to enhance jailbreak defense.}
\label{framework}
\end{figure*}

\section{Preliminary}
In this section, we formulate the existing unlearning defense paradigm and our proposed dynamic jailbreak path unlearning paradigm.
\subsection{Existing Unlearning Defense}

Denote a large language model parameterized by $\theta$, the goal of existing unlearning defense is to update the model parameters to erase specific harmful knowledge while maintaining general capabilities. Specifically, given a forget set $\mathcal{D}_f = \{(x_f, y_f)\}$ containing direct harmful queries and corresponding prohibited responses, and a retain set $\mathcal{D}_r = \{(x_r, y_r)\}$ containing general knowledge samples, the optimization objective of the traditional unlearning defense can be formulated as:
\begin{equation}
\mathcal{L}(\theta) = \mathcal{L}_h(\mathcal{D}_f) + \lambda \mathcal{L}_u(\mathcal{D}_r),
\end{equation}
where $\lambda$ balances the trade-off between safety and utility. Typically, $\mathcal{L}_{\text{u}}$  minimizes the negative log-likelihood on $\mathcal{D}_r$ to preserve general capabilities, while $\mathcal{L}_{\text{h}}$ suppresses harmful memory of LLM either by Gradient Ascent on the harmful pairs $(x_f, y_f)$ or by minimizing the likelihood of a predefined refusal target $y_{refusal}$ (e.g., ``I cannot assist'') given the harmful query $x_f$.

\subsection{Problem Formulation}
We identify that the persistence of dynamic jailbreak paths constitutes the fundamental gap between existing static unlearning methods and robust jailbreak defense. Consequently, we define our research paradigm as follows: given a jailbreak path $\Phi(x)$ triggered by an adversarial input $x$ within $\mathcal{M}_\theta$, the goal is to rectifies the dynamic trajectories converging to harmful outputs. Considering the adaptive nature of attack strategies, we formulate this process as a bi-level optimization problem:
\begin{equation}
\arg\min_{\theta} \max_{\pi} \mathbb{E}_{x \sim \pi} [\mathcal{L}_{\text{p}}(\Phi(x))]  + \lambda \mathcal{L}_{\text{u}}(\mathcal{D}_r).
\end{equation}
In this paradigm, the inner maximization simulates an attacker seeking an optimal strategy $\pi$ to generate jailbreak samples $x$ that maximally activate the model's currently vulnerable paths. The outer minimization represents the defender updating parameters $\theta$ to suppress these dynamically exposed pathways via a path constraint loss $\mathcal{L}_{\text{p}}$, while simultaneously maintaining general capability using $\mathcal{L}_{\text{u}}$. Through this formulation, we transform defense from passive static knowledge deletion into active path blocking, ensuring that even under an evolved attack policy $\pi'$, the induced internal path $\Phi(x')$ fails to reach the harmful semantic terminal.

\section{Methodology}
In this section, we detail JPU, a unified unlearning framework designed to rectify dynamic jailbreak paths towards safety anchors by mining on-policy adversarial samples. As illustrated in Figure \ref{framework}, JPU comprises an Iterative three steps. The first step evolves to mine the most vulnerable adversarial samples for current model. The second step uses these samples to retrace jailbreak path from the intermediate layers to harmful sink nodes in deep layer. In the final step, JPU executes a constraint jailbreak path rectification. 
\subsection{On-Policy Attack Buffer Mining}

Aforementioned empirical analysis attributes the failure of existing unlearning defenses to their inability to rectify diverse jailbreak paths. This necessitates an evolving supply of adversarial data to effectively track and rectify adaptive trajectories.

Drawing inspiration from evolutionary adversarial strategies~\cite{liuautodan,doumbouya2024h4rm3l}, which demonstrate that attack probes must evolve alongside the model to expose the latent vulnerabilities, we propose that the unlearning data for JPU must be dynamically adapt to the model's current state. To this end, we actively maintain an adaptive jailbreak buffer $\mathcal{B}$ to simulate evolving and dynamic attacks. Formally, each entry in $\mathcal{B}$ is structured as a quadruple:
\begin{equation}
    e = (T, Q, J, H),
\end{equation}
where $T$ represents the jailbreak template, $Q$ and $J$ denotes harmful query and complete jailbreak prompt, respectively, and $\mathcal{H}$ records the mutation history. Specifically, we initialize $\mathcal{B}$ with 200 held-out templates and strategies instantiated on the AdvB-Short dataset~\cite{chao2025jailbreaking} (see Appendix~\ref{sec:appendix_buffer} for construction details). 

In each iteration $t$, we lazily sample a batch of seed prompts from $\mathcal{B}$ and compute their respective refusal loss on the current model $\mathcal{M}_t$ regarding a predefined target $y_{ref}$ (e.g.,``I cannot assist''). This loss serves as a proxy for attack success, reflecting the model's current vulnerabilities. We then filter samples using a threshold of $0.5$. Samples exceeding a loss of $0.5$ are retained as parents, as they fail to generate the desired refusal and thereby reveal critical vulnerabilities. Conversely, lower-loss ones indicate successful defense and are discarded to improve buffer efficiency.

Subsequently, we randomly apply homologous mutations $H$ to harmful queries $q$ of parents to generate offspring, expanding the vulnerability search space. Finally, the aggregated on-policy batch $\mathcal{B}_t$, consisting of both parents and offspring, is then passed to the path identification stage.

\subsection{Jailbreak Path identification}

In this step, we aim to identify the jailbreak paths by capturing the critical information trajectory propagated on the on-policy batch. Inspired by Inter-layer Gradient Integration (IGI)~\cite{li2025cross}, which offers axiomatic attribution for locating trajectories through the model's FFN layers, it is intuitive to directly employ it in JPU. However, due to the high computational cost arising from the integral approximation required for the diverse and complete model outputs in their specific tasks, it is prohibitive for JPU's online training efficiency.

To address this challenge, we strategically target a specific harmful semantic sink node of the model output instead of the entire output distribution and employ a first-order Taylor approximation to reduce the computational complexity from $O(|\mathcal{V}|)$ to $O(1)$. Specifically, we first anchor the path identification to the logits of the token ``Sure'' at the last layer. This is grounded in the inherent mechanism of adversarial attacks where an affirmative prefix serves as the standard junction convergence for successful jailbreaks~\cite{zou2023universal}. Starting from this semantic sink, we then retrace the information trajectory back to the FFN layers, which are widely recognized as the key units for knowledge storage and semantic processing in Transformers, by quantifying the contribution of each neuron $i$ at layer $l$ using the flow score, derived as the first-order Taylor proxy of IGI:
\begin{equation}
\text{Flow}^{(l,i)}_{\text{jb}} = |W^{(l,i)}| \cdot |A^{(l,i)} \odot \nabla A^{(l,i)}|,
\end{equation}
where $A_i$, $W_i$, and $\nabla A_i$ denote the activation, weight, and gradient, respectively. 

Additionally, to avoid impairing general functions, we compute the differential flow by subtracting the flow associated with a reference utility batch. Finally, we generate a binary sparse mask $m_t$ by selecting the top-$p\%$ connections that exhibit high jailbreak flow but low utility flow.

\subsection{Constraint Path Rectification}
Given the jailbreak paths embedded in $m_t$, a naive defense strategy would be to simply prune them. However, such modification merely suppresses specific activations and fails to guide the model toward generating constructive safety responses and poses risks of performance degradation, preventing convergence to an optimal defensive state.

In contrast, JPU designs an novel unlearning paradigm by simultaneously imposing a refusal behavior constraint $\mathcal{L}_{\text{h}}$, a safety representation alignment $\mathcal{L}_{\text{s}}$, and a utility preservation constraint $\mathcal{L}_{\text{u}}$. Specifically, $\mathcal{L}_{\text{h}}$ enforces the model to generate standard refusal responses (e.g., ``I cannot fulfill this request''), effectively blocking harmful outputs at the token level. Simultaneously, $\mathcal{L}_{\text{s}}$ minimizes the distance between the current representation $H_{\text{curr}}$ and the centroid of a reference safety anchor $H_{\text{safe}}$ in $\mathcal{M}_t$, thereby explicitly pulling the dynamic jailbreak path back into the safe semantic space. Finally, JPU incorporates a standard Negative Log-Likelihood loss on a utility dataset $\mathcal{D}_r$ to preserve general performance. The algorithm formalization of JPU is provided in Appendix~\ref{alg:JPU}. The complete optimization objective for iteration $t$ is formulated as follows:
\begin{equation}
\begin{split}
\mathcal{L}_{\text{total}} &= \underbrace{\mathbb{E}_{x \in \mathcal{B}_t} [\mathcal{L}_{\text{h}}(x) + \beta \mathcal{L}_{\text{s}}(H_{\text{curr}}, H_{\text{safe}})]}_{\text{Jailbreak Path Rectification on } m_t } \\
&+ \underbrace{\lambda \mathbb{E}_{x \in \mathcal{D}_{\text{r}}} [\mathcal{L}_{\text{u}}(x)]}_{\text{Global Utility Preservation}},
\end{split}
\end{equation}
where $\beta$ balances the explicit refusal behavior with internal representation alignment, while $\lambda$ regulates the trade-off between unlearning plasticity and general utility stability. Through joint unlearning paradigm, JPU achieves a soft update by effectively reprogramming the adversarial circuitry rather than destroying it, allowing the model to converge to an optimal point where robust jailbreak resistance and general utility coexist.

\begin{table*}
\centering
\renewcommand{\arraystretch}{0.95}
\resizebox{1\textwidth}{!}{
\begin{tabular}{lccccccccc}
\toprule
\multirow{2.5}{*}{\textbf{Methods}} & \multicolumn{2}{c}{\textbf{AIM}} & \multicolumn{2}{c}{\textbf{GCG}} & \multicolumn{2}{c}{\textbf{AutoDAN}} & {\textbf{Decoding}} & \textbf{\textsc{MIX-JAIL}} \\
\cmidrule(lr){2-3} \cmidrule(lr){4-5} \cmidrule(lr){6-7} \cmidrule(lr){8-8} \cmidrule(lr){9-9}
& AdvB  & AdvE & AdvB & AdvE & AdvB & AdvE & MaliciousInstruct & AdvB-Short \\
\midrule
\multicolumn{9}{c}{\textit{Llama2-7B-Chat}} \\
\midrule
Base model       & 3.27 & 10.79 & 11.54 & 4.08 & 20.77 & 27.10 & 19.00 & 21.15 \\
\midrule
RSFT             & 0.38 & 0.48  & 2.31  & 0.96 & 8.85  & 16.07 & 9.00  & 15.20 \\
Eraser           & 0.77 & 8.15  & 4.62  & 1.44 & 9.23  & 17.27 & 7.00  & 14.76 \\
Safe Unlearning  & 0.58 & 0.72  & 4.42  & 1.92 & 6.92  & 13.67 & 8.00  & 13.45 \\
Circuit Breaker  & 0.38 & 0.72  & 4.81  & 2.16 & 7.12  & 13.19 & 10.00 & 11.15 \\
CKU              & 0.19 & 0.48  & 4.23  & 1.68 & 6.54  & 12.71 & 7.00  & 8.15 \\
\rowcolor{defgreen} 
\textbf{JPU (Ours)} & \textbf{0.00} & \textbf{0.01} & \textbf{0.58} & \textbf{0.24} & \textbf{3.46} & \textbf{11.59} & \textbf{1.00} & \textbf{4.44} \\
\midrule
\multicolumn{9}{c}{\textit{Llama3-8B-Instruct}} \\
\midrule
Base model       & 3.08 & 9.83 & 9.04 & 3.60 & 18.65 & 24.46 & 17.00 & 16.43 \\
\midrule
RSFT             & 0.38 & 0.24 & 1.92 & 0.96 & 6.54  & 13.91 & 7.00  & 15.10 \\
Eraser           & 0.38 & 6.95 & 3.46 & 1.44 & 7.88  & 15.11 & 8.00  & 15.90 \\
Safe Unlearning  & 0.58 & 0.72 & 3.27 & 1.68 & 7.12  & 10.79 & 7.00  & 15.65 \\
Circuit Breaker  & 0.38 & 0.72 & 3.65 & 1.92 & 7.50  & 11.51 & 8.00  & 14.80 \\
CKU              &\textbf{0.00} & 0.24 & 2.69 & 1.20 & 5.96 & \textbf{9.83} & 6.00 & 14.10 \\
\rowcolor{defgreen} 
\textbf{JPU (Ours)} & \textbf{0.00} & \textbf{0.00} & \textbf{0.96} & \textbf{0.48} & \textbf{5.77} & 11.15 & \textbf{4.00} & \textbf{12.20} \\
\bottomrule
\end{tabular}}
\caption{Comparison of JPU with baselines on Jailbreak attack success rate (ASR). The results show that our method outperforms previous baselines on nearly all attack scenarios, achieving the best resistance against jailbreak.}
\label{tab:main_results}
\end{table*}

\section{Experiment}

In this section, we perform comprehensive evaluations and analysis to evaluate the performance of our proposed jailbreak defense method JPU.

\subsection{Setup}

\noindent \textbf{Data} \space To evaluate jailbreak defense performance, we adopt four jailbreak datasets following compared method CKU~\cite{shi2025safety}: AdvBench~\cite{zou2023universal}, AdvExtent~\cite{lu2024eraser}, MaliciousInstruct~\cite{huangcatastrophic}, and AdvB-Short, a refined 50 samples subset of AdvBench introduced by ~\citet{chao2025jailbreaking}. For general capability assessment, we utilize widely employed benchmarks, including MT-Bench~\cite{zheng2023judging}, CommonsenseQA~\cite{talmor2019commonsenseqa}, HellaSwag~\cite{zellers2019hellaswag}, RTE~\cite{wang2018glue}, WinoGrande~\cite{sakaguchi2021winogrande}, and OpenBookQA~\cite{mihaylov2018can}. Furthermore, to evaluate the exaggerated safety behaviours of models, we employ the XSTest dataset~\cite{rottger2024xstest}.

\noindent \textbf{Victim Models} \space In our experiment, we select two representative open-source safety aligned LLMs, Llama-2-7B-Chat~\cite{touvron2023llama} and Llama-3-8B-Instruct~\cite{dubey2024llama}. 

\noindent \textbf{Metrics} \space For defense evaluation, we employ the Attack Success Rate (ASR) follow the CKU. For general capability evaluation, we conduct a unified assessment using the \texttt{llm-eval}\footnote{\url{https://github.com/EleutherAI/lm-evaluation-harness}} framework.

\noindent \textbf{Baselines} \space We compare JPU against unlearning-based defense baselines: Eraser~\cite{lu2024eraser}, Safe Unlearning~\cite{zhang2024safe}, and CKU. And safety alignment methods: RSFT~\cite{deng2023attack} and Circuit Breaker~\cite{zou2024improving}. The details are provided in Appendix~\ref{sec:appendix_baselines}.

\noindent \textbf{Attack Methods} \space We employ four jailbreak methods utilized in CKU: AIM ~\cite{lu2024eraser}, AutoDAN ~\cite{liuautodan}, GCG ~\cite{zou2023universal}, and Generation Exploitation Attack ~\cite{huangcatastrophic}. For more robust evaluation, we construct \textbf{MIX-JAIL}, a dynamic jailbreak dataset containing 70,000 jailbreak prompts derived from six distinct jailbreak methods. The attack settings and construction details are outlined in Appendix~\ref{sec:appendix_mixjailbreak}.

\noindent \textbf{Setup of JPU} \space
The jailbreak buffer $\mathcal{B}$ in JPU covers three predefined attack types, comprising five distinct jailbreak methods. Additional experimental details are provided in Appendix~\ref{appendix_implementation}.

\begin{table*}
\centering
\tiny
\renewcommand{\arraystretch}{0.95}
\resizebox{\textwidth}{!}{%
\begin{tabular}{l|c|cccc|c}
\toprule
\textbf{Methods} & \textbf{MT Bench} & \textbf{RTE} & \textbf{Op QA} & \textbf{HellaSwag} & \textbf{Co QA} & \textbf{XSTest (FRR)} \\
\midrule
\multicolumn{7}{c}{\textit{LLama2-7B-Chat}} \\
\midrule
Base model & 6.31 & 71.08 & 33.60 & 57.40 & 58.67 & 28.80 \\
\midrule
RSFT & 5.97 & 70.38 & 33.40 & 56.62 & 57.40 & 36.42 \\
Eraser & 5.84 & 70.86 & 33.20 & 57.22 & 58.61 & 33.33 \\
Safe Unlearning & 6.22 & 71.02 & 33.40 & 57.26 & 58.55 & 27.78 \\
Circuit Break & \textbf{6.25} & 71.04 & \textbf{33.60} & \textbf{57.36} & 58.65 & 29.78 \\
CKU & 6.24 & \textbf{71.08} & 33.40 & 57.34 & \textbf{59.01} & 25.56 \\
\textbf{JPU (Ours)} & 6.24  & \textbf{71.08} & \textbf{33.60} & \textbf{57.36} & 58.92 & \textbf{22.40} \\
\midrule
\multicolumn{7}{c}{\textit{LLama3-8B-Instruct}} \\
\midrule
Base model & 8.10 & 67.41 & 33.40 & 57.82 & 75.80 & 3.20 \\
\midrule
RSFT & 7.51 & 66.04 & 32.60 & 56.73 & 74.85 & 6.00 \\
Eraser & 7.86 & 66.94 & 33.00 & 57.04 & 75.48 & 5.22 \\
Safe Unlearning & 7.82 & 67.25 & 32.80 & 57.28 & 75.26 & 5.44 \\
Circuit Break & \textbf{8.06} & 67.26 & \textbf{33.60} & 57.59 & \textbf{75.70} & 5.56 \\
CKU & 7.96 & 67.32 & 33.20 & 57.62 & \textbf{75.70} & 5.11 \\
\textbf{JPU (Ours)} & 7.98 & \textbf{67.39} & 33.20 & \textbf{57.68} & \textbf{75.70} & \textbf{1.60} \\
\bottomrule
\end{tabular}%
}
\caption{Comparison of JPU with baselines on general utility. The evaluation metric for MT-Bench is the average score across two turns and the except NLP tasks is accuracy. The \textbf{ XSTest (FRR) } reports the False Refusal Rate on benign prompts (lower is better). The results show that JPU preserves model's utility and reduces the over-refuse.}
\label{tab:utility_results}
\end{table*}

\subsection{Main Results}

\noindent \textbf{Defense Effectiveness} \space As detailed in Table~\ref{tab:main_results}, JPU consistently outperforms baselines, achieving the lowest ASR in nearly all settings. This validates the efficacy of our proposed jailbreak path unlearning paradigm. For instance, JPU surpassess the best method, CKU, by imporving ASR nearly \(\mathbf{\times 2}\) times when employ  against Llama-2-7B-Chat. Moreover, it is notable that JPU achieves defense performance only utilizing the small subset of randomly selected queries from AdvB-Short for optimization, suggesting that targeting jailbreak paths within LLMs offers superior defense generalization compared to existing knowledge unlearning. 

\begin{figure}[t]
\centering
\includegraphics[width=\linewidth]{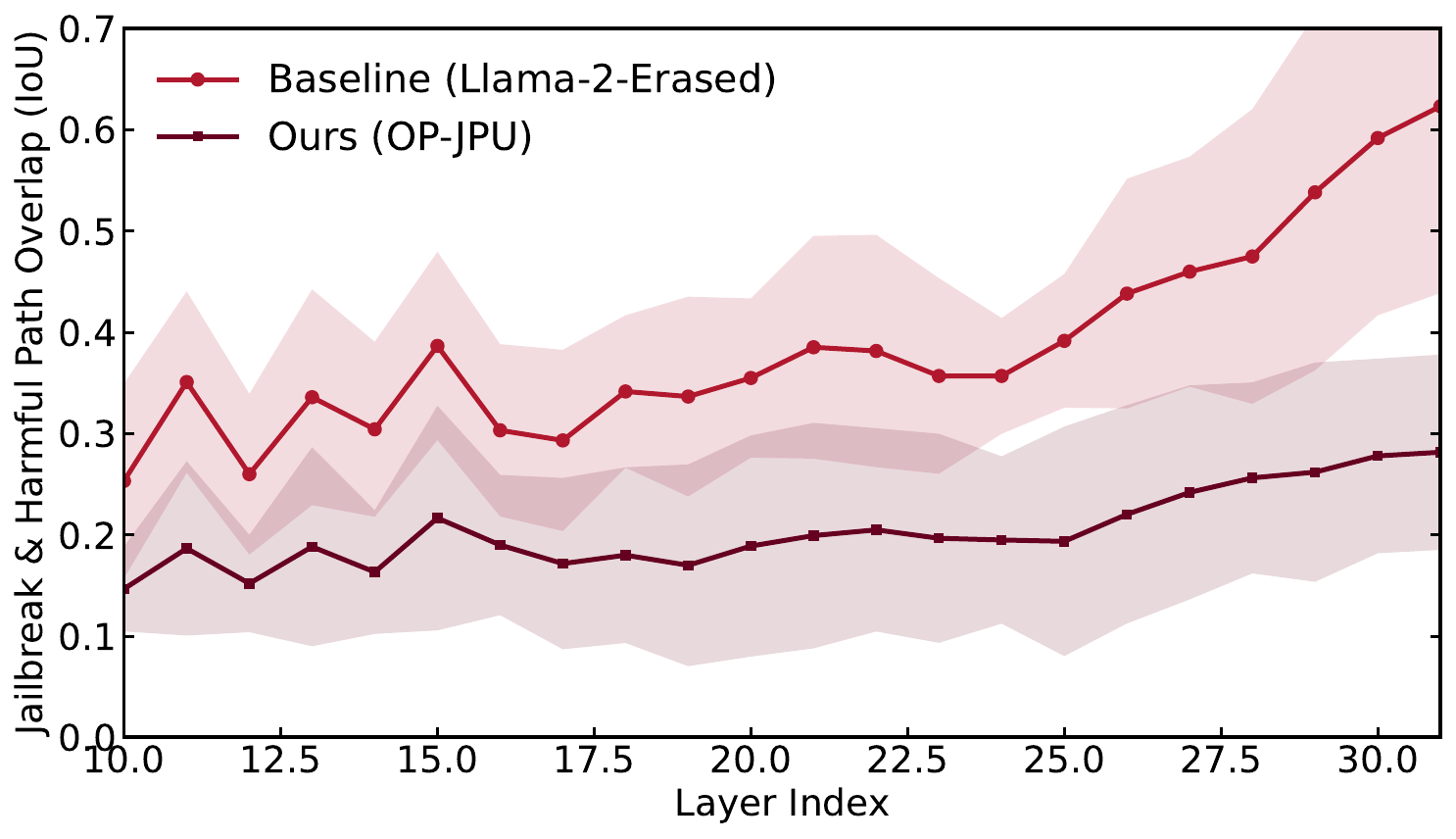}
\caption{Comparison of underlying adversarial circuitry with the baseline (identified by IGI within LLMs). Shaded regions denote the variability across different attack methods, while solid lines represent the averaged layer-wise IOU. The results reveal that our method, JPU, rectifies these jailbreak paths more effectively. }
\label{main_fig_path}
\end{figure}

To verify that JPU indeed rectifies the underlying adversarial circuitry rather than target output patterns, we  follow the same methodology in our empirical analysis to track the overlap of identified harmful and jailbreak path. As illustrated in Figure~\ref{main_fig_path}, JPU demonstrates a significant and consistent reduction in path overlap across critical layers for attack methods within MIX-JAIL compared to the baseline~\cite{lu2024eraser}. This quantitatively confirms that JPU successfully mitigating dynamic jailbreak paths at the structural level.

\begin{figure}[t] 
\centering
\includegraphics[width=\linewidth]{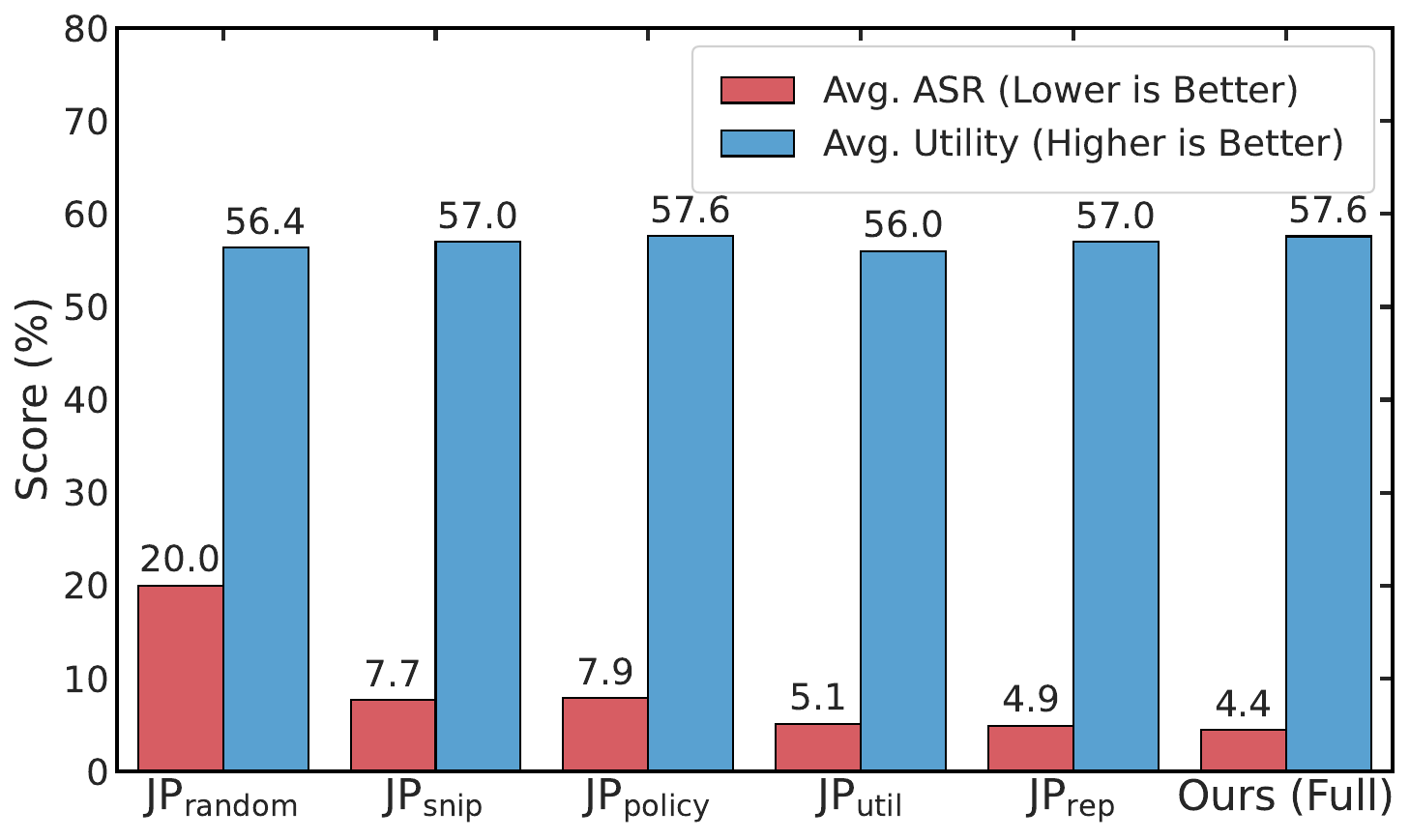} 
\caption{Ablation experiments illustrating the impact of different components of JPU. Each part of JPU plays a vital role in enhancing jailbreak resistance ability and maintaining general ability. }
\label{ablation_fig_components}
\end{figure}

\noindent \textbf{General Effectiveness.} \space  Table~\ref{tab:utility_results} represents JPU consistently matches model on NLP tasks, preserving utility. Moreover, regarding the False Refusal Rate on XSTest, JPU outperforms baselines and even reduces the over-sensitivity of the original model. For instance, JPU reduces FRR by \(\mathbf{\times 2}\) times on Llama3, indicating that it successfully disentangles adversarial inputs from benign semantics rather than relying on indiscriminate refusal.

\subsection{Ablation Studies}
We conduct a series of experiments to comprehensively evaluate the effectiveness of components and key hyperparameters within JPU. For fair comparison, all experiments are performed on Llama-2-7B-Chat using the \textsc{MIX-JAIL} dataset.

\noindent \textbf{Impact of Components.} \space 
To evaluate the contribution of each JPU component, we construct five variants: (1) JP\textsubscript{random}: JPU randomly masks an equal number of parameters; (2) JP\textsubscript{policy}: JPU utilizes a fixed buffer without adversarial mining;  (3) JP\textsubscript{snip}: JPU forms a path using the top-$p$ neurons ranked by SNIP; (4) JP\textsubscript{rep}: JPU without safety anchor alignment; (5) JP\textsubscript{util}: JPU without utility preservation constraint. As illustrated in Figure~\ref{ablation_fig_components}, removing any component degrades either defense robustness or general utility. Specifically, replacing jailbreak path localization with random masking JP\textsubscript{random} substantially weakens defense performance, causing ASR to approach that of the base model, which confirms the effectiveness of our path localization.
Moreover, JP\textsubscript{snip} slightly outperforms CKU which adopts SNIP localization, indicating that the remaining components further rectifies latent attack pathways beyond parameter selection alone.

\begin{table}[t]
\centering
\resizebox{\columnwidth}{!}{
\begin{tabular}{l|ccc|c}
\toprule
\textbf{Model Name} & \textbf{GPTFuzzer} & \textbf{ReNeLLM} & \textbf{Jailbroken} & \textbf{Avg.} \\
\midrule
JPU\textsubscript{shallow} & 10.51\% & 36.67\% & 4.17\% & 23.05\% \\
JPU\textsubscript{middle} & 1.22\% & 24.92\% & 15.97\% & 13.08\% \\
JPU\textsubscript{last} & 3.58\% & 17.66\% & 18.75\% & 10.78\% \\
JPU & \textbf{0.08\%} & \textbf{12.51\%} & \textbf{3.19\%} & \textbf{6.54\%} \\
\bottomrule
\end{tabular}
}
\caption{Results of attack success rate (ASR) on JPU with different unlearning layer selection strategies. The results show that the original strategy (from intermediate to the last layers) is the most effective for identifying the comprehensive jailbreak path.}
\label{tab:albation_layer_selection}
\end{table}


\noindent \textbf{Impact of Layer Selection} \space 
To assess JPU's layer selection strategy for identifying path, we compare three selection variants, including: JPU\textsubscript{shallow} (layers 0--10), JPU\textsubscript{middle} (layers 11--20), and JPU\textsubscript{last} (the last four layers), with  original strategy from intermediate to the last layer. 

Table~\ref{tab:albation_layer_selection} presents the results against attack methods sampled from \textsc{MIX-JAIL}. Our analysis yields two findings. First, intervening at the shallow and middle layers leads to a significant degradation in defense, revealing  jailbreak semantics and paths have not ye fully formed in these layers. Second, JPU\textsubscript{last} achieves closest performance to the full method, showing that disrupting the harmful aggregation of jailbreak paths and guiding them toward safe regions plays a vital role in JPU. However, it neglects signals propagated from intermediate layers, thereby narrowing the scope of defense and resulting in compromised performance.

\begin{figure}[t]
\centering
\includegraphics[width=0.9\linewidth]{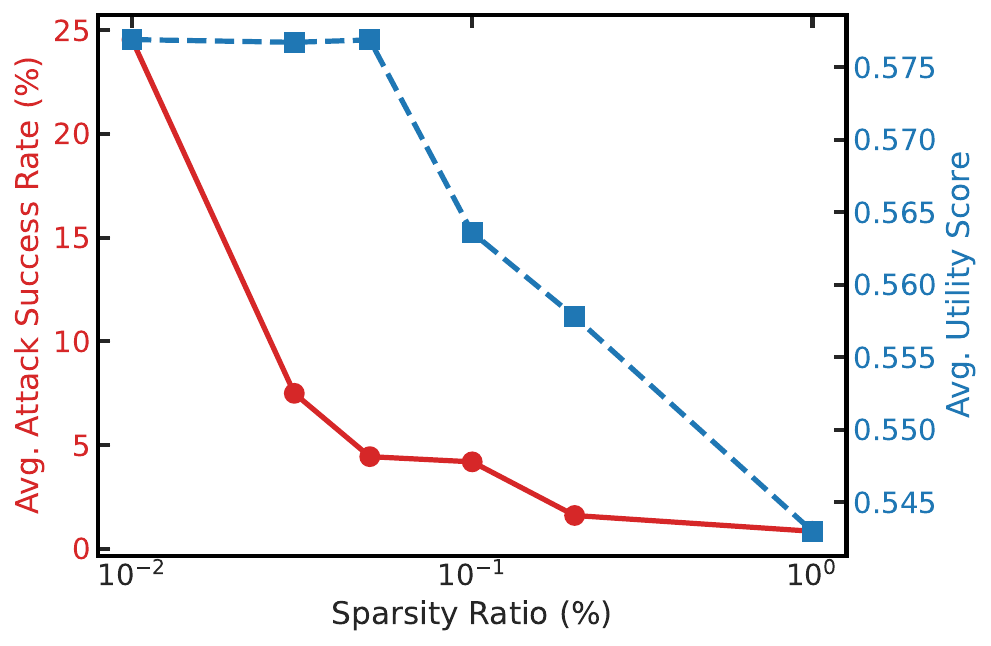}
\caption{This figure illustrates how neuron sparsity of the jailbreak path influences JPU achieves balance between model utility and jailbreak resistance.}
\label{ablation_fig_sparsity}
\end{figure}

\noindent \textbf{Impact of Path Sparsity} \space 
To examine the impact of path sparsity (defined as the top-$p\%$ masked neurons) on JPU, we evaluate its performance on Mix-JailExpert across $p \in \{0.01, 0.03, 0.05, 0.1, 0.2, 1.0\}$. Figure~\ref{ablation_fig_sparsity} highlights a utility-safety trade-off: extremely small $p$ compromises defense effectiveness, while excessively large $p$ degrades utility. Therefore, we adopt $p=0.05$ as our standard configuration, as it strikes the optimal balance in our experiments.

\subsection{Necessity of On-Policy Mining Strategy}
To empirically validate the essential necessity of the on-policy mining for capturing dynamic attack features, we conduct ablation studies comparing training strategies: Static (Off-Policy) and Dynamic (On-Policy). Specifically, the static strategy relies on a fixed buffer without on-policy attack buffer mining component, while the dynamic strategy corresponds to the full JPU framework.

\begin{figure}[t]
\centering
\includegraphics[width=\linewidth]{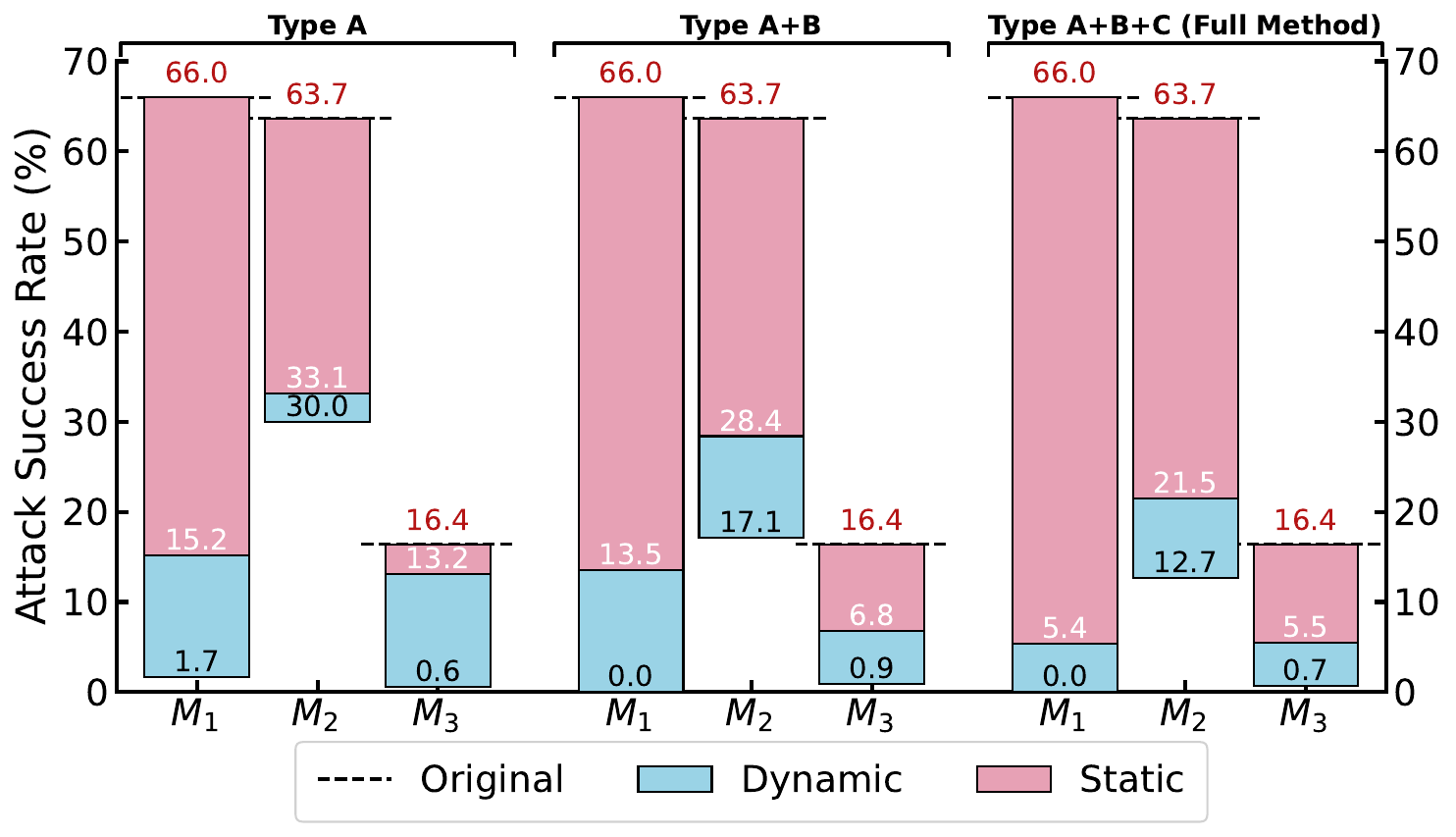} 
\caption{The results are visualized as overlapping bars, where the static and dynamic results are plotted against the original model's baseline (Dashed Line). A lower bar height indicates a more effective reduction in ASR. }
\label{ablation_fig_onpolicy}
\end{figure}

To ensure comprehensive robustness against variations in data scale and diversity, we evaluate both strategies under three progressive buffer initialization settings: Type $\in$ \{A, A+B, A+B+C (full)\}. We assess the overall defense performance against three representative attack methods ($M_1$-$M_3$) from \textsc{MIX-JAIL} dataset on Llama-2-7B-Chat. More information is detailed in \ref{on_policy_ablation_details}. As illustrated in Figure~\ref{ablation_fig_onpolicy}, the dynamic strategy consistently achieves a substantially lower ASR than the static baseline across all buffer settings. Even with the full diversity buffer, the static variant remains inferior to the on-policy strategy, indicating that simply scaling static data is insufficient for addressing dynamic threats. Instead, achieving an effective defense performance depends on on-policy adversarial mining, which actively identifies and mitigates evolving vulnerabilities that static datasets fail to capture.


\section{Conclusion}
In this paper, we introduce JPU, a novel framework designed to rectify dynamic jailbreak paths via on-policy rectification. Our research reveals a critical limitation in existing static unlearning defenses: adaptive attacks can activate the non-erased parameters in intermediate layers and eventually align with harmful semantics paths to trigger unsafe outputs. Our experimental results demonstrate that JPU not only achieves superior defense performance against diverse and dynamic jailbreak attacks but also effectively preserves the general utility of LLMs. Moreover, post-hoc mechanism analysis confirms that our method successfully rectifies dynamic jailbreak paths and redirects these malicious paths toward safety regions. We hope that our work provides valuable insights into the dynamics of jailbreak attacks and inspires future research on robust unlearning-based defenses.

\section{Limitations}

The current implementation of our Attack Buffer within JPU is primarily designed to encompass prompt-based injection strategies. While JPU exhibits strong generalization capabilities against decoding-based attacks, exploring the integration of attack vectors beyond the prompt space—such as parameter-level manipulation or optimization-based attacks—could potentially yield even more robust defense performance. Integrating a broader spectrum of adversarial methodologies remains a promising direction to provide deeper insights and guidance for the design of future safety alignment strategies.


\appendix

\section{Details of Baselines}
\label{sec:appendix_baselines}

In this section, we provide detailed descriptions of the baseline defense methods employed in our comparative experiments. These baselines encompass two primary categories: Unlearning-based defenses and Representation-based defenses.

\subsection{Unlearning-based Defenses}
\begin{itemize}
    \item \textbf{Eraser}~\cite{lu2024eraser}: \space
    It operates on the principle of Gradient Ascent. Specifically, Eraser maximizes the loss on harmful queries to decrease the likelihood of generating hazardous tokens, while simultaneously employing a retain loss on general domain datasets to prevent catastrophic forgetting. The optimization objective is to push the model's distribution away from the harmful target sequences, erasing the knowledge required to answer malicious queries.
    
    \item \textbf{Safe Unlearning}~\cite{zhang2024safe}: \space
    It treats safety training as a negative preference optimization problem. It constructs pairs of harmful query and response and minimizes the probability of generating the harmful response while steering the model towards a safe distribution or a specific refusal target. It emphasizes minimizing the side effects of unlearning, aiming to maintain the model's helpfulness on benign instructions while removing unsafe behaviors.
    
    \item \textbf{CKU (Confused Knowledge Unlearning)}~\cite{shi2025safety}: \space
    It represents a state-of-the-art approach to knowledge unlearning. Specifically, CKU first identifies safety-critical neurons in the multi-layer perceptron (MLP) components of LLMs that are strongly associated with safeguard knowledge. It then performs targeted unlearning by selectively modifying except these identified neurons, aiming to suppress unsafe behaviors while preserving the model’s safety.
\end{itemize}

\subsection{Representation-based Defenses}

\begin{itemize}
    \item \textbf{RSFT}~\cite{deng2023attack}: \space
    It implements an iterative fine-tuning strategy designed to harden Large Language Models (LLMs) against adversarial prompts. By simulating dynamic interactions between the attacker and the target model, RSFT progressively identifies safety vulnerabilities and refines the model's refusal boundaries, significantly enhancing its resilience to complex, multi-turn harmful instruction attacks.
    
    \item \textbf{Circuit Breaker (Representation Engineering)}~\cite{zou2024improving}: \space
    This is a representation-level defense that utilizes ``circuit breakers'' to reroute harmful internal activations. By intervening at the hidden state level rather than the output level, it provides a generalized defense against diverse attack vectors while preserving the model’s original utility on benign tasks.
\end{itemize}

\section{Preparation of Experiments}
\label{appendix_implementation}

\subsection{Construction of \textbf{MIX-JAIL}}
\label{sec:appendix_mixjailbreak}

In this section, we present the construction details of our proposed challenging jailbreak dataset, \textbf{MIX-JAIL}. 
Specifically, we first collect six representative jailbreak attack methods that have emerged recently, covering diverse attack paradigms. 
These include Type A attacks such as CodeChameleon (4 jailbreak prompts); 
Type B attacks including GPTFuzzer (481 jailbreak prompts), ReNeLLM (475 jailbreak prompts), AutoDAN-Turbo (9 jailbreak prompts), and JailBroken (24 jailbreak prompts); 
as well as Type C attacks such as AmpleGCG (488 adversarial suffixes).

We then apply all collected jailbreak prompts and adversarial suffixes to the AdvB-Short dataset. 
After data filtering and separation, we construct a mixed jailbreak dataset consisting of approximately 70,000 jailbreak instances, which we refer to as \textbf{MIX-JAIL}.

\subsection{Details of Attack Buffer Construction}
\label{sec:appendix_buffer}

The adversarial buffer $\mathcal{B}$ employed in our method integrates a diverse spectrum of existing prompt-based jailbreak methodologies. To ensure comprehensive coverage of the attack surface, we categorize these methods into three distinct types:
\begin{itemize}
\item \textbf{Type A (Code-based Encapsulation)}: Attacks that utilize coding scenarios or nested formats to bypass safety filters, represented by CodeChameleon~\cite{lv2024codechameleon} and RmcBench~\cite{chen2024rmcbench}.
\item \textbf{Type B (Mutation-based Optimization)}: Methods that employ fuzzing or evolutionary strategies to generate complex, evolving attack templates. Representative methods include GPTFuzzer~\cite{yu2023gptfuzzer}, ReNeLLM~\cite{ding2024wolf}, and AutoDAN-Turbo~\cite{liu2024autodan}.
\item \textbf{Type C (Adversarial Suffix Optimization)}: Approaches that optimize fixed adversarial character sequences (suffixes) to induce affirmative responses, such as GCG~\cite{zou2023universal} and AmpleGCG~\cite{liao2024amplegcg}.
\end{itemize}

\noindent \textbf{Buffer Instantiation.} 
The collection process for our 200-sample buffer follows a protocol similar to the construction of the \textsc{MIX-JAIL} benchmark. Specifically, we initialize the buffer by curating distinct jailbreak templates and mutation strategies from the aforementioned categories. We collect 66 templates for Type A (from CodeChameleon and RmcBench), 66 seeds for Type B (aggregated from GPTFuzzer, ReNeLLM, and AutoDAN-Turbo), and 66 suffixes for Type C (from AmpleGCG). Finally, these curated seeds are instantiated by applying them to harmful intents sampled from the AdvB-Short dataset, thereby forming the complete adversarial queries for the buffer $\mathcal{B}$.

\subsection{Hyperparameter Settings}
Regarding the critical coefficients in our joint unlearning objective, we empirically set the representation alignment weight $\beta = 4.0$ and the utility preservation weight $\lambda = 1.0$. Specifically, the value of $\beta=4.0$ is chosen to prioritize the rectification of internal adversarial circuitry, enforcing a strong alignment between the model's hidden states and the safety direction. The utility coefficient $\lambda=1.0$ is selected to maintain a balanced trade-off, providing sufficient regularization to preserve general language capabilities without overly constraining the plasticity required for unlearning jailbreak paths. These hyperparameters were determined based on a grid search on a held-out validation set.

\section{On-Policy Mining Ablation Details}
\label{on_policy_ablation_details}
We selected \texttt{codeChameleon}, \texttt{AutoDAN-Turbo}, and \texttt{GPTFuzzer} from the MIX-JAIL attack dataset to represent models $M_1$, $M_2$, $M_3$, respectively. Subsequently, we conducted ablation experiments on the On-Policy module.

\begin{algorithm*}[t]
\caption{Training Procedure of JPU}
\label{alg:JPU}
\begin{algorithmic}[1]
\Require 
    Initial LLM $\mathcal{M}_\theta$; 
    Initial Jailbreak Buffer $\mathcal{B}$ (containing templates \& mutation strategies);
    Utility Dataset $\mathcal{D}_r$; 
    Reference Safety Anchor $H_{\text{safe}}$;
    Refusal Target $y_{\text{ref}}$ (e.g., ``I cannot assist'');
    Semantic Sink Node $y_{\text{sink}}$ (e.g., token ``Sure'').
\renewcommand{\algorithmicrequire}{ \textbf{Input:}} 
    Jailbreak threshold $\tau$; 
    Masking ratio $p$; 
    Loss weights $\beta, \lambda$; 
    Learning rate $\eta$.
\Ensure Defended Model $\mathcal{M}^*_\theta$.

\While{not converged}
    \State \textbf{// Stage 1: On-Policy Attack Buffer Mining}
    \State Sample a batch of seed prompts $S_{\text{seed}}$ from Buffer $\mathcal{B}$.
    \State Compute refusal loss $\mathcal{L}_{\text{ref}}(x)$ on current $\mathcal{M}_\theta$ for $x \in S_{\text{seed}}$ w.r.t $y_{\text{ref}}$.
    \State Identify \textbf{Parents} (High-loss samples): $\mathcal{P} \leftarrow \{x \in S_{\text{seed}} \mid \mathcal{L}_{\text{ref}}(x) > \tau\}$.
    \State Discard low-loss samples $S_{\text{seed}} \setminus \mathcal{P}$.
    \State Generate \textbf{Offspring} $\mathcal{O}$ by applying homologous mutations to $\mathcal{P}$.
    \State Aggregate On-Policy Adversarial Batch: $\mathcal{B}_t \leftarrow \mathcal{P} \cup \mathcal{O}$.
    \State Update Buffer $\mathcal{B}$ with new effective samples from $\mathcal{B}_t$.

    \State \textbf{// Stage 2: Jailbreak Path Identification}
    \State For $\mathcal{B}_t$, target the semantic sink node $y_{\text{sink}}$ at the last layer.
    \State Compute \textbf{Jailbreak Flow Score} via first-order Taylor approximation:
    \begin{equation*}
        \text{Flow}^{(l,i)}_{\text{jb}} = |W^{(l,i)}| \cdot |A^{(l,i)} \odot \nabla A^{(l,i)}|
    \end{equation*}
    \State Compute \textbf{Utility Flow Score} $\text{Flow}_{\text{util}}$ on a batch from $\mathcal{D}_r$.
    \State Calculate Differential Flow: $S^{(l,i)} = \text{Flow}^{(l,i)}_{\text{jb}} - \text{Flow}^{(l,i)}_{\text{util}}$.
    \State Generate \textbf{Sparse Mask} $\mathcal{M}_t$: Set top-$p\%$ connections with highest $S^{(l,i)}$ to 1, others to 0.

    \State \textbf{// Stage 3: Constraint Path Rectification}
    \State Compute Rectification Loss on $\mathcal{B}_t$:
    \begin{equation*}
        \mathcal{L}_{\text{rect}} = \mathbb{E}_{x \in \mathcal{B}_t} [\mathcal{L}_{\text{h}}(x) + \beta \mathcal{L}_{\text{s}}(H_{\text{curr}}, H_{\text{safe}})]
    \end{equation*}
    \State Compute Utility Loss on $\mathcal{D}_r$: $\mathcal{L}_{\text{util}} = \mathbb{E}_{x \in \mathcal{D}_r} [\mathcal{L}_{\text{u}}(x)]$.
    \State Compute Gradients and Apply Soft Update:
    \begin{equation*}
        \theta \leftarrow \theta - \eta \left( (\nabla_\theta \mathcal{L}_{\text{rect}} \odot \mathcal{M}_t) + \lambda \nabla_\theta \mathcal{L}_{\text{util}} \right)
    \end{equation*}
\EndWhile
\Return $\mathcal{M}_\theta$
\end{algorithmic}
\end{algorithm*}

\end{document}